\documentclass[prl,aps,twocolumn]{revtex4}
\usepackage{graphicx}
\begin{document}
\author{A. T. Costa, Jr.}  
\altaffiliation[Permanent address: ]{Departamento de Ci\^encias Exatas, 
Universidade Federal de Lavras, 37200-000 Lavras, MG, Brazil}
\email{a.costa@qubit.org}
\affiliation{Centre for Quantum Computation -
University of Oxford, Parks Road, OX1 3PU, Oxford, United Kingdom}
\author{R. B. Muniz} 
\altaffiliation[Permanent address: ]{Instituto de F\'\i sica, Universidade 
Federal Fluminense, 24210-340 Niter\'oi, RJ, Brazil}
\email{rbechara@duey.ps.uci.edu}
\affiliation{Department of Physics and Astronomy -
University of California, Irvine, CA, United States}

\title{Spin filtering by ferromagnetic nanowires}

\begin{abstract}

We show that electrical current flowing through nanowires made of
ferromagnetic disordered alloys can become highly spin polarized.

\end{abstract}


\maketitle

The increasing ability to control spin-dependent phenomena in
condensed matter physics is opening exciting possibilities for the
electronic industry. The search for efficient spin filtering systems
in particular is very important for spin-based electronics and quantum
computation. Spin polarized electrical current for example may be
required to operate qubits in some proposed device schemes for quantum
computations \cite{Awschalom,divincenzo,Burkard}.

Recently, a clever and promising technique has been devised for
fabricating very thin metallic nanowires, with diameters smaller than
10nm \cite{HDai}. It uses suspended carbon nanotubes as substrates for
deposition by electron-beam evaporation of several metals, including
Fe and Ni. By initially coating the carbon nanotube with a small
amount of Ti (with nominal thicknesses ranging from 1-2nm) other
metals can stick more easily to the substrate, leading to the
formation of nice continuous nanowires up to tens of microns long.

The purpose of this letter is to show that electrical current flowing
through nanowires of ferromagnetic disordered alloys can become highly
spin polarized. Therefore, wires with such characteristics may be
useful as spin injectors in electronic devices. We begin by presenting
and discussing the main physical mechanisms that are responsible for
the spin filtering behavior of these systems. Then we perform model
calculations to illustrate how effective those mechanisms can be, and
finally discuss our results and the guidelines they provide for
choosing the nanowire composition that may maximize the spin filtering
effect.

It is well known that disorder in metallic systems may lead to
localization. Depending on the nature and degree of disorder,
localized states may appear near the top and bottom of the conduction
bands of three dimensional metallic systems, with a mobility edge
separating localized from extended states. Non-interacting disordered
electronic systems with lower dimensionality in the thermodynamic
limit have localized states only. Strictly speaking they are
insulators at zero temperature \cite{Lee,Thouless}. In fact,
localization effects can be very effective in reducing the conductance
of nanowires \cite{McKinnon}. They can make the average conductance
$\bar{g}$ fall off exponentially with the wire length $\ell$ for
sufficiently long wires. The localization length $\Lambda$ is
determined by the asymptotic decaying rate of $\bar{g}$, being shorter
the faster $\bar{g}$ decreases. Nikoli\'c and MacKinnon have made a
detailed study of the electrical conductance in non-magnetic
disordered nanowires, employing a single-band tight-binding model to
describe the electronic states \cite{McKinnon}. The problem involves
basically four characteristic lengths: $\ell$, $\Lambda$, the wire
width $w$, and the electronic mean free path $\lambda$. In their work,
they illustrated the occurrence of different transport regimes
according to existing relations between $\ell$, $\lambda$, and
$\Lambda$. Transport is quasi ballistic, when $\ell$ is comparable
with $\lambda$, mesoscopic when $\lambda < \ell < \Lambda$, and
strongly localized when $\Lambda < \ell$. The quantities $\lambda$,
$\Lambda$, and $\bar{g}$ are all functions of energy, and the
conductance may change regime as the energy varies.  Actually, in the
presence of bulk disorder they found that $\bar{g}$ decays faster as a
function of $\ell$ for energies close to the band edges in the strong
localization regime.

It is noteworthy that the scattering of carriers in metallic
ferromagnets is generally spin dependent, even when the scattering
potentials do not depend upon spin. This is mainly due to the
densities of states around the Fermi energy being different for
majority and minority spin carriers. In ferromagnetic transition
metals, for example, the spin polarization of the sp-electrons is
relatively small compared with that of the d-electrons. Both sp and
d-electrons participate in electrical conductance, but the d-electrons
are less mobile because they have a larger effective mass.
Nevertheless, even assuming the current is predominantly carried by
sp-electrons in such systems, sp-d hybridization may lead to distinct
conductances for up and down spins. The reason, as rightfully argued
by Mott, is that electrical resistance is proportional not only to the
density of scattering centers, but also to the number of available
states where electrons can scatter into. Thus, the existence of
unoccupied d states at the Fermi energy ($E_F$) in transition metals
acts as a trap for the sp electrons, because sp-d hybridization allows
them to be scattered into the available d-states. Since the densities of
available d-states at $E_F$ differ for up and down spins in
ferromagnetic transition metals, it follows that the electronic mean
free paths of majority and minority spin carriers are usually not
equal in such systems. Furthermore, the atomic potential fluctuations
experienced by d electrons in transition metal alloys are often much
larger than those felt by s electrons. As a consequence, s and d
electrons are differently affected by disorder, and we will show that
the localization lengths for majority and minority spin electrons may
be also rather different in ferromagnetic nanowires made of transition
metal alloys. In such systems $\lambda$, and $\Lambda$, are both
energy and spin dependent quantities.

An additional relevant length scale for discussing spin dependent
transport in ferromagnets is the mean free path associated with spin
flip scattering ($\lambda_f$). In metallic systems $\lambda_f$ is
usually much longer than $\lambda$. For nanowires with $\ell <
\lambda_f$, the conductances for up and down spins are independent,
hence $\bar{g} = \bar{g}_{\uparrow} + \bar{g}_{\downarrow}$, where
$\bar{g}_{\sigma}$ is the average conductance for electrons with spin
$\sigma$. The up- and down-spin channels in this case behave as
resistors in parallel. Since both $\bar{g}_{\uparrow}$ and
$\bar{g}_{\downarrow}$ depend upon energy, they may decay as functions
of $\ell$ with different rates, particularly in strong ferromagnets
where the majority d-bands are completely filled.  With dissimilar
localization lengths for up and down spin carriers, the polarization
of the electrical current may thus increase very rapidly with $\ell$.

To illustrate how effective these mechanisms can be, we perform model
calculations for ferromagnetic nanowires made of metallic disordered
alloys. For numerical simplicity we consider two-dimensional wires of
finite lengths, sandwiched by perfect leads. Experimentally the wire
thickness is not constant, and this is simulated by assuming it
fluctuates randomly by $\pm \delta w$ around an average width
$\bar{w}$. In order to allow spin dependent Mott scattering mechanism
to take place, the nanowire electronic structure is described by a
simple s-d model Hamiltonian
 
\begin{equation}
H = \sum_{ij\sigma} \sum_{\mu\nu} h_{ij}^{\mu\nu}
    a^{\dagger}_{i\mu\sigma}a_{j\nu\sigma} +
    \sum_{i\mu} U_{i}^{\mu} n_{i\mu\uparrow}n_{i\mu\downarrow}\, ,
\label{hamilt}
\end{equation}
where $a^{\dagger}_{i\mu\sigma}$ is the creation operator for an
electron with spin $\sigma$ in orbital $\mu$ on site $i$, and
$n_{i\mu\sigma}$ is the corresponding occupation number. The sp- and
d-orbital sets in transition metals are represented here by just two s
orbitals labeled by $\mu=1,2$, respectively. The on-site matrix
elements $h_{ii}^{\mu\nu} = \epsilon_i^\mu \delta_{\mu\nu} +
(1-\delta_{\mu\nu}) \gamma_i$, where $\epsilon_{i}^{\mu}$ are atomic
energies, and $\gamma_i$ characterizes the hybridization between the s
and d bands. Our system consists of a ferromagnetic disordered alloy,
where sites are occupied with a certain probability either by magnetic
or non-magnetic metal atoms.  We neglect the effect of disorder in the
hopping integrals, and consider them as being non zero between nearest
neighbor sites only. Thus, for nearest neighbor sites $i \neq j$,
$h_{ij}^{\mu\nu} = -t^{\mu\mu}\delta_{\mu\nu}$, where $t^{ss}$ and
$t^{dd}$ symbolize the transfer integrals for s and d electrons,
respectively. $U_{i}^{\mu}$ represents the Coulomb interaction between
electrons located on the same site and orbital, and we further assume
it takes place only when they occupy the d-orbitals of the magnetic
metal atoms.

In principle all the parameters involved here can be estimated from
the band structures or renormalized atomic potentials of the
constituent metals. However, given the model nature of our calculation
we have selected representative values for the parameters, rather than
attempting to adjust them to fit a specific system. For example, we
choose our energy unit such that $t^{ss}=1$, and take $t^{dd}=0.2$,
and $\gamma_i=0.9$ independent of the lattice sites. This
approximately reproduces the typical bandwidths of transition
metals. By placing the band structures of the constituents on the same
absolute scale, and aligning the corresponding Fermi energies, one may
also estimate the energy levels $\epsilon_{H(I)}^{s,d}$,
characterizing the host $(H)$ and impurity $(I)$ metals. We shall
discuss alloys based on ferromagnetic transition metals such as Fe, Co
or Ni, with non-magnetic transition metal impurities of the left side
of those elements in the periodic table. We thus set $\epsilon_H^s=0$
as our energy origin, and choose $\epsilon_H^d = 0$,
$\epsilon_I^s=0.03$, and $\epsilon_I^d = 0.6$ or $0.9$, as a reasonable set
of values for describing representative systems in the scope of such
simple a model.

We treat the electron interaction within the Hartree-Fock
approximation, thus reducing the on-site interaction term to
$-\frac{1}{2}\Delta_i^\mu \delta_{\mu\nu}\sigma$, where $\Delta_i^\mu$
represent the exchange splittings for the s and d orbitals, and
$\sigma=\pm 1$ for $\uparrow$ and $\downarrow$ spins,
respectively. With the simple form of interaction assumed, only
$\Delta_H^d\neq0$, corresponding to a molecular field acting solely on
the d orbitals of the magnetic-metal host atoms.

Since our wire may be viewed as a sequence of atomic chains of finite
sizes $\bar{w} \pm \delta w$, it is convenient to label the atomic
site positions by a pair of indices ($l$,$r_l$) representing the line
$l$ the atom belongs to, and its position $r_l$ along that line.  The
conductance in the spin channel $\sigma$ is calculated by the Kubo
formula \cite{Mathon}

\begin{equation}
g_{\sigma} (E) = \frac{4e^2}{h}{\rm Re}{\rm
Tr}\lbrack\tilde{G}^{\sigma}_{00}t_{01}\tilde{G}^{\sigma}_{11} t_{10}-
t_{01}\tilde{G}^{\sigma}_{10}t_{01}\tilde{G}^{\sigma}_{10}\rbrack\, ,
\label{kubo}
\end{equation}
where $0$ and $1$ symbolize any two adjacent line indices (the choice
is arbitrary due to current conservation), and $t_{01}$ is the
tight-binding hopping matrix between such lines.
$\tilde{G}^{\sigma}_{lm}=\frac{1}{2i}(G^{\sigma -}_{lm} - G^{\sigma
+}_{l,m})$, where $G^{\sigma\pm}_{l,m}$ are matrices representing the
advanced and retarded one-electron propagators for particles with spin
$\sigma$, evaluated at energy $E$, connecting lines $l$ and $m$. ReTr
stands for the real part of the trace over all orbitals and line
sites, and $e^2/h$ is the quantum conductance.  

The required one-electron Green functions were calculated by the
method described in reference \cite{Mathon}. Firstly, the
surface Green functions of the semi-infinite perfect leads are
generated by a well established technique. \cite{Lopez-And} Then, the
Dyson equation is employed recursively to built the corrugated
disordered wire parts atop the leads to the left and right of lines
$0$ and $1$, respectively. Finally, those wire parts are reconnected
by turning on the electron hopping $t_{01}$ between them. In our
calculations the impurity sites are randomly chosen for a given alloy
concentration, and configurational averages of $g_{\sigma}$ are
performed over 4,000 different samples.

Before presenting our results for ferromagnetic systems, it is
instructive to study the conductance of non-magnetic disordered wires
with this simple s-d model, particularly the behavior of ${\bar g}$
as a function of $\ell$ for different values of energy. Thus, we start
by setting $\Delta_H^d = 0$, and consider wires with $\bar{w} = 10$
atoms, $\delta w = \pm1$ atom, sandwiched by leads of width
$w_L=\bar{w}$ (leads of larger widths play no significant role on the
main points we wish to address).  Figure 1 shows localization lengths
calculated as functions of energy for different impurity potentials
and concentrations. Here, it is statistically more appropriate to
define $\bar{g} = \exp\lbrack\langle\ln g\rangle\rbrack$, where
$\langle f\rangle$ denotes the configurational average of $f$ \cite
{McKinnon}. $\Lambda$ is determined by fitting the asymptotic
exponential decaying rate of $\bar{g}$ with $\ell$; more precisely,
$\Lambda^{-1} = -\partial \bar{g}/\partial\ell$ for $\ell\gg 1$.  The
localization length clearly depends upon the impurity potential
strength relative to the host, i.e., on $\Delta\epsilon^d =
\epsilon_I^d - \epsilon_H^d$.  It is also evident that $\Lambda^{-1}$
increases as energy approaches the d-bands region (delimited here by
$\approx\epsilon^d\pm0.8$ for non-hybridized bands), where conduction
electrons have more states to be scattered into. States in this energy
range are more affected by impurity concentration changes because
$\mid\Delta\epsilon^d\mid \gg \mid\Delta\epsilon^s\mid$.  The most
striking feature, however, is the very pronounced maximum that appears
in $\Lambda^{-1}$ near the top of the d band, for reasonably large
positive values of $\Delta\epsilon^d$.  The maximum becomes bigger and
broader with increasing impurity concentration, and corresponds to a
considerably faster decaying of $\bar{g}$ with $\ell$, as illustrated
in the inset of figure 1(a). It arises from an interesting combination
of disorder, hybridization and screening effects in the nanowire.  To
understand its physical nature we recall that hybridization mixes the
s- and d-conducting channels, and d-states are more directly affected
by disorder in transition metal alloys. Screening in such systems is
also very effective, and can cause significant changes in the local
density of states (LDOS) around the impurity sites. It is well known
that resonances and virtual bound states may appear near $E_F$,
depending on the relative strength of the impurity potential. At
sufficiently low temperatures, the conduction electrons are scattered
predominantly into available states near the scattering centers.
Therefore, impurity resonant states at $E_F$ may strongly influence
electrical conductance, especially in disordered metallic systems. We
notice that the energy position and some features of the
$\Lambda^{-1}$ maximum are closely related to the impurity potential
strength. This is illustrated in the inset of figure 1(b), where we
compare the LDOS on a single impurity site, and its averaged value
over the impurity neighboring sites, with $\Lambda^{-1}$; all
calculated in the same energy range for $\Delta\epsilon^d=0.6$, but
smaller values of $x=0.05$, and $\gamma=0.2$, in order to highlight
the virtual bound state structure in the LDOS. The peaks in
$\Lambda^{-1}$ and in the LDOS are clearly correlated in this case, as
expected.

\begin{figure}
\includegraphics[scale=0.45]{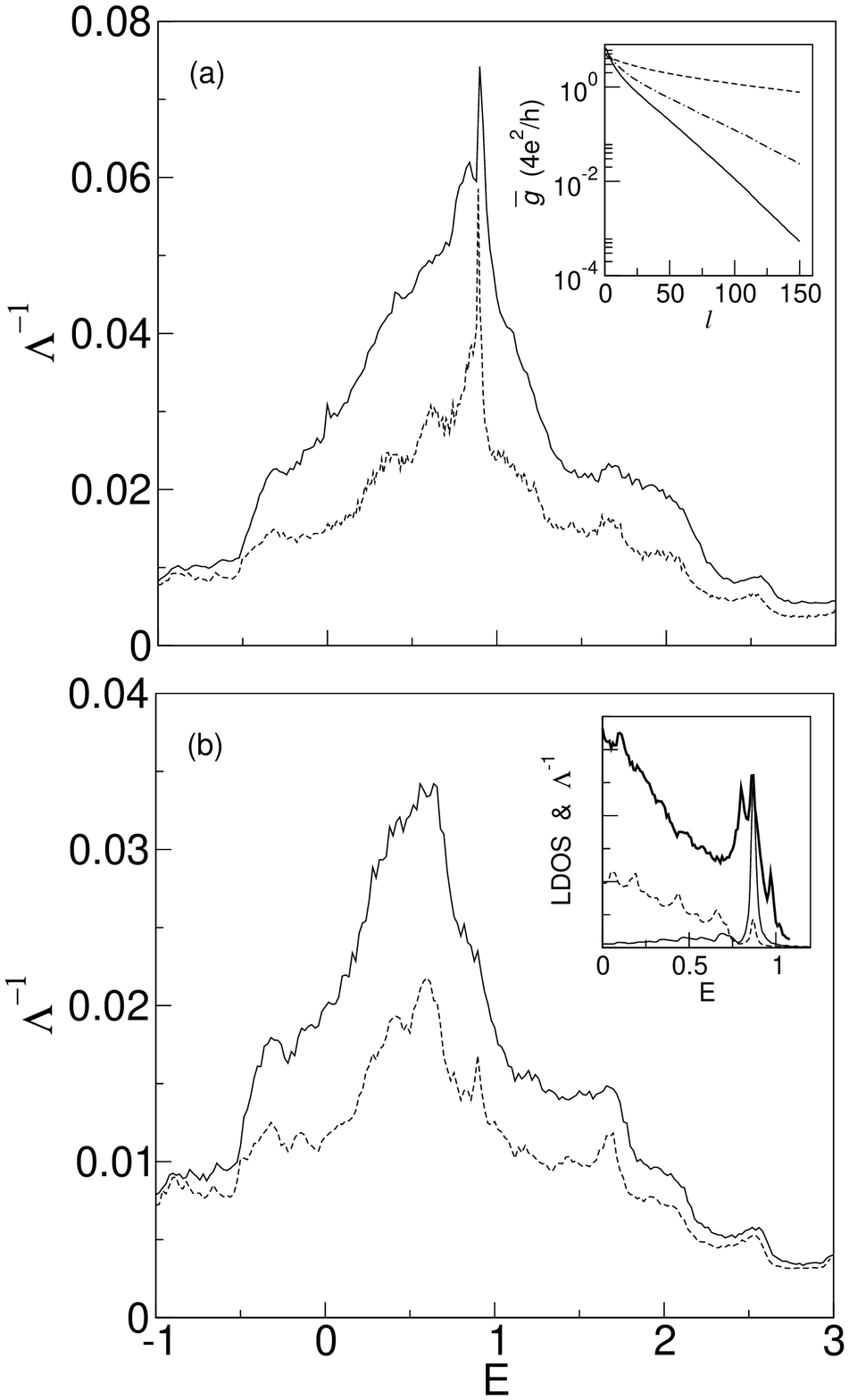}
\vskip -3 cm
\caption{Inverse localization lengths calculated for a non-magnetic
disordered wire as functions of energy for different impurities
($\Delta\epsilon^d=0.9$ (a); $\Delta\epsilon^d=0.6$ (b)), and
concentrations ($x=0.1$ (dashed line); $x=0.2$ (solid line)).  The
inset in (a) shows the average conductance (calculated for
$\Delta\epsilon^d=0.9$ and $x=0.2$) plotted against the wire length
$\ell$ for distinct values of energy: $E_F=2.4$ (dashed line),
$E_F=1.2$ (dot-dashed line), and $E_F=0.8$ (solid line). Lengths and
energies are measured in units of the lattice spacing and s-hopping
integral, respectively.  The inset in (b) shows the LDOS (in arbitrary
units) on an impurity site (dashed line), and averaged over its
surrounding sites (thin solid line), together with $\Lambda^{-1}$
(thick solid line), all calculated in the same energy range for
$\Delta\epsilon^d=0.6$, but smaller values of $x=0.05$, and
$\gamma=0.2$.}
\label{fig1}
\end{figure}

In fact, screening and disorder effects in ferromagnetic transition
metal alloys can be highly spin dependent. In some Ni based alloys,
screening is done largely by down-spin electrons, whereas up-spins
normally play a significant part in Fe based alloys.
Ni$_{1-x}$Ti$_x$, for instance, is a classical example where the
introduction of a Ti impurity leads to an up-spin bound state being
pushed completely above $E_F$ \cite{Edwards}. The characters of up-
and down-spin conducting states can certainly be altered by varying
the alloy constituents and composition. To illustrate how appropriate
combinations of elements may lead to very large spin filtering effects,
we finally consider a ferromagnetic disordered nanowire attached to
two non-magnetic metallic leads.  We assume a typical value of
exchange splitting $\Delta_H^d = 0.5$ on the d orbitals of the
magnetic host atoms, and choose $E_F= 1.2$, corresponding to a strong
ferromagnet with its up-spin d band completely filled. Our results for
$\bar{g}_{\uparrow}$ and $\bar{g}_{\downarrow}$, calculated as
functions of $\ell$, are shown in figure 2. They depict
$\bar{g}_{\downarrow}$ decreasing much faster than
$\bar{g}_{\uparrow}$, showing that the localization length of up-spin
electrons in this case is much larger than that of the
down-spins. Such a remarkable behavior is basically due to the large s
character of the up-spin conduction states, contrasting with the much
less mobile and predominant d character of the down-spin ones.  As a
consequence, the percentage fraction of up-spin conductance increases
rapidly with the wire length. This is represented in the inset of
figure 2, where $P=\bar{g}_{\uparrow}/\bar{g}$ is plotted against
$\ell$ for two different values of $E_F$, corresponding to a strong
and weak ferromagnet, respectively. One immediately sees that very
high spin polarized electrical currents can be achieved with
relatively short ferromagnetic disordered nanowires, just a few
hundred atoms long. This makes them excellent candidates for being
used as spin injectors in nanoelectronic devices.

\begin{figure}
\vskip -0.8 cm
\includegraphics[scale=0.32]{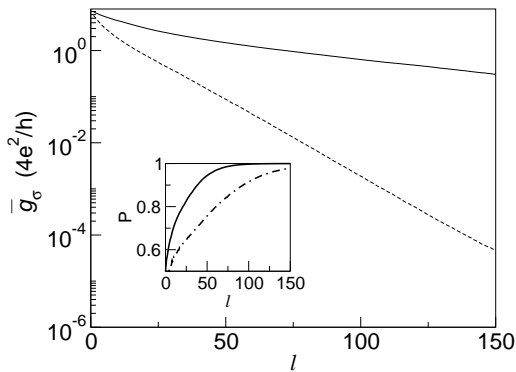} 
\vskip -0.5 cm
\caption{Average conductances of majority- (solid line) and
minority-spin electrons (dashed line), calculated as functions of wire
length $\ell$, for ferromagnetic disordered nanowires. The
calculations were performed for $\Delta^d=0.9$, $E_F=1.2$ and $x=0.2$
(see text). Lengths and energies are measured in units of the lattice
spacing and s-hopping integral, respectively.  The inset shows the
percentage fraction of up-spin conductance
$P=\bar{g}_{\uparrow}/\bar{g}$ plotted against $\ell$ for $E_F=1.2$
(solid line) and $E_F=0.35$ (dot-dashed line), corresponding to a
strong and weak ferromagnet, respectively}
\label{fig2}
\end{figure}

In summary we have shown that localization lengths of up- and
down-spin electrons may be rather different in ferromagnetic
disordered nanowires. As a result, electrical currents flowing through
such structures may acquire a high degree of spin polarization. The
localization lengths for up and down spins are determined mainly by
the character of the electronic states around the Fermi energy. By
judiciously choosing the nanowire alloy constituents and
concentration, one can certainly modify such characters, and perhaps
even control those localization lengths. There is a large ground for
experimenting with such systems. Presently, the nanowires reported in
reference \cite{HDai} are composites of a carbon nanotube in the core,
covered by a relatively small amount of Ti, and a second metal on the
outside. Nevertheless, careful heat treatments, and simultaneous
deposition of different metals, may possibly be employed to fabricate
nanowires of ferromagnetic disordered alloys. The most obvious choices
to start with would be NiTi and FeTi alloys. The Ti impurity potential
strength is relatively stronger in Ni alloys, but the magnetization of
FeTi remains finite for larger concentrations of Ti. Other transition
metal combinations, may deserve to be examined.  The fact that both Ni
and Fe separately form continuous nanowires, suggests that NiFe alloys
with virtually any concentration may also do, thus broadening the
possibilities for exploring different spin dependent characteristics
of the conducting states.  NiCr alloys, with the Fermi level sitting
on an up-spin virtual bound state, seems an interesting system
too. The spin filtering mechanism reported here relies on localization
and hybridization effects. Most importantly, on the existence of
relatively large contrast between the conduction state characters of
up and down spins in the nanowires. We hope our findings will
stimulate further investigation on these systems.

\begin{acknowledgments}
We thank S. L. A. de Queiroz and D. L. Mills for valuable comments.
Financial support from CNPq, FAPERJ and FAPEMIG (Brazil), is
gratefully acknowledged.
\end{acknowledgments}

\printfigures

\end{document}